\documentclass{article}

\usepackage{arxiv}

\usepackage[utf8]{inputenc} 
\usepackage[T1]{fontenc}    
\usepackage{hyperref}       
\usepackage{url}            
\usepackage{booktabs}       
\usepackage{amsfonts}       
\usepackage{nicefrac}       
\usepackage{microtype}      
\usepackage{lipsum}

\title{PASS: \textit{De novo} assembler for short peptide sequences}

\author{
   Ren\'e L.~Warren\\
  Genome Sciences Centre, BC Cancer\\
  Vancouver, BC, V5Z 4S6, Canada \\
  \texttt{rwarren@bcgsc.ca} \\
}

\begin{document}
\maketitle

\begin{abstract}
The ability to characterize proteins at sequence-level resolution is vital to biological research. Currently, the leading method for protein sequencing is by liquid chromatography mass spectrometry (LC-MS) whereas proteins are reduced to their constituent peptides by enzymatic digest and subsequently analyzed on an LC-MS instrument. The short peptide sequences that result from this analysis are used to characterize the original protein content of the sample. Here we present PASS, a \textit{de novo} assembler for short peptide sequences that can be used to reconstruct large portions of protein targets, a step that can facilitate downstream sample characterization efforts. We show how, with adequate peptide sequence coverage and little-to-no additional sequence processing, PASS reconstructs protein sequences into relatively large ($\geq$100 amino acid) contigs having high (93.1 - 99.1\%) sequence identity to reference antibody light and heavy chain proteins.

\textbf{Availability:} PASS is released under the GNU General Public License Version 3 (GPLv3) and is publicly available from \url{https://github.com/warrenlr/PASS}.
\end{abstract}


\section{Introduction}
In the past two decades, we have witnessed the rapid progress and growth in the development and world-wide adoption of DNA/RNA sequencing technologies\cite{erwin}. The rise of each new commercial technology, often accompanied with massive sequence throughput and orders of magnitude lower per-base cost, has created opportunities for bioinformatics tool development, to help assemble, analyze and characterize millions of nucleic acid sequences with various length profiles and error modalities. In contrast, advances in peptide sequencing have not followed the same trajectory\cite{floyd}, and peptide sequencing by liquid chromatography mass spectrometry (LC-MS) is still the leading and most robust technology for protein identification and characterization. As part of the LC-MS protocol, proteins are enzymatically digested to their peptide components, and analyzed by LC-MS, resulting in short ($\leq$50 residues) peptide sequences that are used to determine the original protein components of the sample using proprietary software.\par
In the early days of the next-generation sequencing revolution, when the nucleic acid sequences produced by the Solexa/Illumina Ltd instrument were still very short ($\leq$25 bp), we developed SSAKE, a \textit{de novo} genome assembler for short sequencing reads capable of complete and near-complete viral (30 kbp) and bacterial (5 Mbp) genome assembly \cite{warren}. SSAKE cycles through short sequence reads stored in a hash table and progressively searches through a prefix tree for extension candidates. The short read assembler has since been adapted to recognize the amino acid alphabet and the new utility, the Protein Assembler with Short Sequences (PASS), was developed to reconstruct \textit{de novo} underlying protein sequences like the ones analyzed by mass spectrometry, solely from short ($\geq$6 residues) and relatively low-coverage peptide sequences.\par
Here, we describe the PASS algorithm and present initial benchmarks on an experimental high-coverage antibody mass spectrometry peptide sequence data. PASS is available from \url{https://github.com/warrenlr/PASS} and provided with the hope that it is useful to the proteomics and scientific community at large.\par

\section{Methods}
\subsection*{Algorithm}
A. Sequence Overlap\par
Short peptide sequences of length l in a single multi-FASTA input file -f are read in memory, populating a hash table keyed by unique sequence with pairing values representing the number of sequence occurrence in the input sequence set. The normalized peptide sequences are sorted by decreasing abundance (number of times the sequence is repeated) to reflect coverage and minimize extension of sequences containing sequencing errors. Peptide sequences with erroneous residues are more likely to be unique in the entire input data set when compared to their error-free counterparts. Sequence assembly is initiated by generating the longest COOH-most word (k-mer) from the unassembled peptide sequence u that is shorter than the sequence length l. Every possible COOH-terminal most k-mers will be generated from u and used in turn for the search until the word length is smaller than a user-defined minimum, m. Meanwhile, all perfectly overlapping sequences will be collected in an array and be further considered for COOH-termini extension once the k-mer search is complete. At the same time, a hash table c will store every amino acid residue along with a coverage count for every position of the overhang (or stretches of amino acids hanging off the seed sequence u).\par
Once the search completes, a consensus sequence is derived from the hash table c, taking the most represented amino acid at each position of the overhang. To be considered for the consensus, each amino acid has to be covered by user-defined -o (set to 2 by default). If there is a tie (two amino acids at a specific position having the same coverage count), the prominent amino acid is below a user-defined ratio r, the coverage -o is too low or the end of the overhang is reached, the consensus extension terminates and the consensus overhang joined to the seed sequence/contig. All overlapping sequences are searched against the newly formed contig sequence and, if found, are removed from the hash table and prefix tree. If they are not part of the consensus, they will be used to seed/extend other contigs, if applicable. If no overlapping peptide sequences match the newly formed contig, the extension is terminated from that end and PASS resumes with a new seed sequence. That prevents infinite looping through low-complexity amino acid sequences. In the former case, the extension resumes using the new [l-m] space to search for joining k-mers.\par
The process of progressively cycling through longer to shorter COOH-most k-mer is repeated after every sequence extension until nothing else can be done on that sequence end. Since only left-most searches are possible with a prefix tree, when all possibilities have been exhausted for the COOH-terminal extension, the complementary strand of the contiguous sequence generated is used to extend the contig on the NH2-end. The prefix tree is used to limit the search space by segregating peptide sequences and their reverse counterparts (if applicable) by their first 5 amino acid at the NH2-termini.\par

There are three ways to control the sequence assembly stringency in PASS. 1) Disallow sequence/contig extension if the coverage is too low (-o). Higher -o values lead to shorter contig lengths, but minimizes sequence misassemblies. 2) Adjust the minimum overlap -m allowed between the seed/contig and short sequences. Higher m values lead to more accurate contigs at the cost of decreased contiguity. 3) Set the minimum amino acid ratio -r to higher values.\par

B. Using a seed sequence\par

If the -s option is set and points to a valid FASTA file, the protein sequences comprised in that file will populate the hash table and be used exclusively as seeds to nucleate contig extensions (they will not be utilized to build the prefix tree). In that scheme, every unique seed will be used in turn to nucleate an extension, using short sequences found in the tree (specified in -f). This feature might be useful if you already have characterized sequences and wish to increase their length, using short peptide sequences. Since short sequences are not used as seeds when -s is set, they will not cluster to one another without a seed sequence file. Also, to speed up the assembly, no embedded sequences (i.e. those aligning to the seed in their entirety) are considered. Only peptide sequences that contribute to extending a seed sequence are noted. When -s is set, the output .contigs file lists all extended seeds, even if it is by a single amino acid. The .singlets output file will only list seeds that could not be extended. Unassembled microsequences will not be captured in the output files.\par

C. \textit{De novo} assembly with a seed sequence\par

The -i option instructs PASS to use target sequences for the sole purpose of recruiting peptide sequences. If set (-i 1), the target sequences will recruit peptide sequences for independent, \textit{de novo} assemblies. This has the advantage of allowing the user to provide, as a target, a large reference sequence (-s) without \textit{a priori} knowledge of variant amino acids. PASS does not constrain the k-mer length derived from a target sequence for interrogating candidate sequences. User-defined target word length values are passed to the algorithm using the -j option. Using larger -j values should help speed up the search when using long peptide sequences, since it will restrict the sequence space accordingly. Note: whereas specificity, speed and RAM memory usage may increase with -j, it may yield more sparse/fragmented assemblies. Systematic experimentation with various -j values are therefore warranted.\par

\subsection*{Data}
The antibody peptide sequence LC-MS/MS data analyzed herein was downloaded from \url{https://www.ncbi.nlm.nih.gov/pmc/articles/PMC4999880/bin/srep31730-s2.zip} and was released as a benchmark dataset for evaluating the ALPS system\cite{tran}. For testing the ability of PASS to assemble peptide sequences \textit{de novo}, we generated a FASTA file for the two antibody data sets, each including a heavy chain and a light chain, and considered short peptide sequences listed in the file peptides\_db.csv exclusively. The human antibody light and heavy chain input files comprised 11,855 and 13,189 peptide sequences, respectively, ranging in size from (minimum length) 6 amino acids to (maximum length) 46 and 43 amino acids, respectively. The WIgG1 antibody light and heavy chain input files comprised 9,211 and 8,072 peptide sequences, respectively, ranging in size from (minimum length) 4 amino acids to (maximum length) 65 and 50 amino acids, in this order. We used a more limited dataset than that used in the ALPS study\cite{tran}, to highlight the general utility of PASS, and not its performance in assembling complete antibody sequences, which cannot be compared fairly to a multi-input pipeline with additional post-assembly sequence alignments steps designed to merge output sequence contigs.\par

\subsection*{Runs}
We ran PASS on the four antibody peptide sequence datasets stated above (version 0.3 with parameters -m 4 -w 10 -o 1 -r 0.51 -t 2) and evaluated resulting output contigs $\geq$100 amino acids in length and with a protein sequence identity $\geq$90\% compared to their respective, intended reference sequences.

\begin{table}[h!]
\hrule \vspace{0.1cm}
\caption{\label{tab:ig}\textit{De novo} peptide assembly of antibody sequences with PASS. Only sequence contigs longer than 100 amino acids are reported.}
\centering
\begin{tabular}{lccc}
\toprule
\textbf{Antibody} & \textbf{ID | Length (aa) | \# of peptides | fold-coverage} & \textbf{Identity (\%)} & \textbf{Coverage (\%)} \\
\midrule
 Human light chain (216aa) &      contig1 | size115 | read4237 | cov642.17 &      96.3 &     48.6  \\
 Human heavy chain (446aa) &      contig2 | size351 | read4855 | cov207.75 &      97.4 &     76.0  \\
 WIgG1 light chain (219aa) &      contig1 | size116 | read774 | cov108.74 &      99.1 &     100.0  \\
  &      contig2 | size120 | read637 | cov85.15 &      98.2 &     -  \\
 WIgG1 heavy chain (441aa) &      contig1 | size323 | read1250 | cov67.18 &      99.1 &     90.7  \\
  &      contig2 | size102 | read409 | cov56.29 &      93.1 &     -  \\
 
\bottomrule
\end{tabular}
\end{table}

\section{Results}
We assembled an external LC-MS/MS peptide sequence dataset corresponding to antibody light and heavy chains, with PASS. Output contigs 100 amino acids and larger are compiled (Table 1). We observed an appreciable assembled contig length (351 and 323 amino acids for the human and WIgG1 heavy chains, respectively), and a high sequence identity (93.1 - 99.1\%) that approaches what was reported using ALPS, a multi-step assembly approach to reconstruct protein sequences from LC-MS data\cite{tran}. Of note, the high fold-coverage of the human light and heavy chain contigs (642 and 207x) and, surprisingly, a lower sequence identity and overall coverage of the reference is observed from those contigs, when compared to the longest ones obtained with the sparser WIgG1 datasets (Table 1). All four assemblies ran fast on a single CPU (Elapsed wall clock time human heavy chain [m:ss] 0:00.64), typically in less than 1 sec on an Intel(R) Xeon(R) CPU E7-8867 v3 @ 2.50GHz system, and used approximately 20MB of RAM per assembly.\par

\section{Discussion and Conclusions}
PASS is a proteomics application for \textit{de novo} assembly of thousands to millions of very short (6 aa) to longer (100 aa) peptide sequences, and beyond. It is derived from the SSAKE genome assembler \cite{warren}, an easy-to-use, robust, reliable and tractable assembly algorithm for very short sequence reads, such as those generated by Illumina Ltd. Algorithms of SSAKE are the core of many genomics applications and their design continues to inspire new-generation assemblers and analysis pipelines \cite{viquas} \cite{novobreak}. Here, we showed how PASS can be used to quickly assemble, \textit{de novo}, thousands of short antibody peptide sequences into sizeable and accurate sequence contigs that can be used for the downstream characterization of a protein target. We expect PASS to have broad applications in proteomics research.\par

\bibliographystyle{unsrt}  


%

\end{document}